\documentclass[runningheads]{llncs}
\usepackage[T1]{fontenc}
\usepackage{graphicx}
\usepackage{color}
\usepackage{hyperref}

\usepackage{amsmath}
\usepackage{amsfonts}
\usepackage{booktabs}
\usepackage{tabularx}
\usepackage{subcaption}
\usepackage{multirow}
\usepackage{orcidlink}
\usepackage{hyperref}

\usepackage{eso-pic}
\usepackage{xcolor}

\renewcommand{\orcidID}[1]{\href{https://orcid.org/#1}{\orcidlink{#1}}}

\begin{document}
\title{LLM-based Source Code Compression via Thresholded Symbol Ranking}
%
\titlerunning{LLM-based Symbol Ranking for Code Compression}
%

\newcommand{\PF}[1]{\textcolor{red}{{\bfseries PF: }#1}}
\newcommand{\AN}[1]{\textcolor{blue}{{\bfseries AN: }#1}}


\author{
Angelo Nardone\inst{1}\orcidID{0009-0006-2068-5934}
\and
Paolo Ferragina\inst{1,2}\orcidID{0000-0003-1353-360X}
}
\authorrunning{A. Nardone and P. Ferragina}
%

\institute{
University of Pisa, Department of Computer Science, Pisa, Italy\\
\email{angelo.nardone@phd.unipi.it} \\
\and
Scuola Superiore Sant’Anna, Pisa, Italy\\
\email{paolo.ferragina@santannapisa.it}
}

\maketitle              

\AddToShipoutPictureFG*{%
  \AtPageLowerLeft{%
    \raisebox{1.5cm}{%
      \makebox[\paperwidth]{%
        \parbox{0.8\paperwidth}{\centering\scriptsize\color{gray}
        This is the version of the paper submitted and then accepted for presentation at the 24th International Conference of the Italian Association for Artificial Intelligence (AIxIA 2026), Lecture Notes in Artificial Intelligence (LNAI), Springer, Perugia, Italy, 6--9 October 2026.}%
      }%
    }%
  }%
}

\begin{abstract}

We study the problem of lossless compression of source code, motivated by the storage demands of large-scale software archives, such as Software Heritage (\url{https://www.softwareheritage.org/}). General-pur\-pose compressors (e.g., \texttt{zstd}, \texttt{bzip2}) offer a good trade-off between compression ratio and speed, but fail to exploit all special regularities inherent in source code. Recent approaches leverage Large Language Models (LLMs) within Shannon's symbol-ranking framework, relying on a scheme in which the predicted rank can grow arbitrarily. While effective at reducing space, this setting incurs significant throughput degradation, and leaves open the question whether it is necessary to explicitly encode all ranks. In this work, we introduce LLM-based compressors deploying two novel symbol-ranking variants that bound predictions to the top-$T$ ranks ($T=1$ or $63$), with out-of-threshold symbols stored as exceptions and compressed jointly with the rank stream via general-purpose compressors. We conduct the first large-scale evaluation of LLM-based source code compression across 30 LLMs, including general-domain, code-specialized, and quantized models. Our $T$-bounded approach outperforms prior LLM-based compressors both in compression ratio (up to 37\% relative improvement) and compression throughput (40\% faster). Compared to general-purpose compressors (e.g., \texttt{zstd}, \texttt{bzip2}), we obtain up to 82\% relative compression gain but at a lower speed, thus offering a new trade-off point in the compression-speed spectrum. We also show that these gains are stronger on source code than on natural language, suggesting an interesting indication, namely that source code exposes regularities captured by LLMs but missed by general-purpose exact-match-based compressors. We conclude by commenting on open problems that offer theoretical and practical avenues of research.

\end{abstract}
\section{Introduction}
\label{sec:introduction}

Data compression is a cornerstone of modern computing, reducing storage, bandwidth, 
and memory pressure while improving cache efficiency~\cite{Navarro:2016book}. 
This is especially critical for lossless compression, which guarantees exact 
reconstruction in domains where fidelity is essential, such as textual data, 
databases, and source code. The need for efficient lossless source-code compression 
is amplified by the rapid growth of software repositories: platforms such as GitHub 
and GitLab now host hundreds of millions of projects, and in 2024 alone GitHub 
reported more than 100 million new projects and nearly 1 billion 
contributions~\cite{github2025newsinsights}.

This growth affects large-scale archival infrastructures such as the 
\emph{Software Heritage Archive (SWH)}, launched by INRIA and UNESCO to collect, 
preserve, and provide universal access to publicly available source code -- an effort 
essential for long-term archival, reproducibility, and for providing a reliable codebase for AI model training. While efficient methods exist for compressing the Merkle DAG underpinning SWH's infrastructure~\cite{fontana:webgraph,morris2003components}, the main challenge lies in compressing the source-code files themselves -- commonly 
referred to as \textit{blobs}. These constitute the overwhelming majority of SWH's 
storage footprint and are currently compressed independently with \texttt{gzip}-based 
tools, a strategy that fails to capture redundancies across files. Several approaches 
have been proposed to address this limitation:
\cite{boffa:compressibility-swh} showed that compressing files jointly via proper 
ordering and large-context-window compressors can partially exploit such redundancies, 
yet at the cost of expensive reorderings and with moderate compression gains.

These limitations motivate new approaches to source-code compression. Since LLMs are effective at modeling sequences, capturing long-range dependencies, and 
predicting upcoming tokens~\cite{deletang2024language}, we investigate their use as 
predictors within Shannon's \emph{symbol-ranking} 
framework~\cite{shannon:prediction-entropy}. Our goal is twofold: to assess whether LLM-based symbol ranking can improve compression ratio while retaining practical throughput, thus bringing such techniques 
closer to deployment in large-scale archives, and to advocate 
for lossless compression as a principled, task-agnostic lens to study and compare 
foundation models.

\smallskip
\noindent\textbf{State of the Art.} Lossless compression via symbol ranking has a 
long history~\cite{shannon:prediction-entropy}. Early implementations include the 
Move-to-Front (MTF) compressor~\cite{bentley:mtf} and Prediction by Partial Matching 
(PPM)~\cite{cleary1997unbounded,cleary1984ppm,ferragina:pearls,lelewer1991streamlining,moffat1990ppm}. 
A major step forward was the Burrows-Wheeler Transform (BWT)~\cite{burrows:wheeler}, refined by Fenwick~\cite{fenwick:block-sorting}, which, with MTF and RLE, underpins widely used compressors (e.g., \texttt{bzip2}~\cite{bzip2manual}).

The connection between language modeling and compression has long been 
established~\cite{mahoney2000fast,schmidhuber1996sequential}, and the rise of neural 
networks has revitalized this perspective. Early neural compressors, such as 
\emph{DeepZip}~\cite{goyal2018deepzip} and \emph{NNCP}~\cite{bellard:nncp}, combined 
RNN/LSTM-based prediction with Arithmetic Coding, achieving interesting compression ratios at very high computational cost. With the advent of 
Transformers~\cite{vaswani2017attention}, LLM-based approaches emerged as strong 
candidates for lossless compression. \emph{LLMZip}~\cite{valmeekam:llmzip} was among the first to 
exploit LLMs for compression, considering both Symbol Ranking, combined with 
\texttt{zlib}, and Arithmetic Coding driven by LLM-predicted probability 
distributions. \emph{FineZip}~\cite{mittu:finezip} explicitly builds on 
\emph{LLMZip}, adding parameter-efficient fine-tuning (LoRA) and dynamic 
context windows; consistently with \emph{LLMZip}, it considers both Symbol Ranking 
with \texttt{bzip2} and Arithmetic Coding, with the latter slightly improving 
compression but being much slower. \emph{AlphaZip}~\cite{narashiman:alphazip} 
focused on smaller models to improve throughput, pairing Symbol Ranking with 
\texttt{gzip} or Brotli. Finally, Del\'etang et al.~\cite{deletang2024language} 
conducted a broader study of foundation models of different sizes -- Chinchilla 
(70B, 7B, 1B), LLaMA~2 (7B), and vanilla Transformers (3.2M, 800K, 200K) -- combined 
with Arithmetic Coding, showing that they can act as effective compressors for diverse data types, including English text, images and videos, while highlighting trade-offs involving model size, context 
length, tokenization, and computational cost.

All these works share a common principle: leveraging a (neural) predictor to assign ranks by 
decreasing predicted likelihood and encoding them via a general-purpose compressor \cite{alakuijala2018brotli,deutsch1996rfc1950,peter1996gzip,bzip2manual}, or 
directly feeding predicted probabilities into Arithmetic Coding \cite{witten1987arithmetic}. In both cases, every 
symbol is assigned a rank or probability, corresponding to Shannon's unbounded 
setting ($T=\infty$) -- no threshold is applied to separate high-confidence predictions 
from rare symbols. While surpassing general-purpose compressors in compression ratio, both 
approaches suffer from very low throughput, preventing their use in real-world 
scenarios, and leave open the question about the impact of $T$ on the final compression ratio.

\smallskip
\noindent\textbf{Our Contribution.} In this work, we investigate LLM-based lossless 
compression for source code, with the following key contributions.

\smallskip
\noindent
\textbf{--} \textit{Novel symbol-ranking pipelines.} We revisit \emph{symbol ranking for text 
compression}, introduced by Shannon in his 1951 work~\cite{shannon:prediction-entropy}, 
and employ LLMs to generate context-driven rankings of the top-$T$ most probable 
symbols. Shannon formalized three prediction schemes: (i) guessing only the most 
likely symbol ($T=1$), with a yes/no outcome and symbol revelation upon error; (ii) 
scanning the full ranked list until the correct symbol is found ($T=\infty$); and 
(iii) a thresholded variant that stops after $T$ incorrect guesses and reveals the 
symbol. We revisit this \emph{thresholded} scheme and propose 
LLM-based variants focusing on two configurations ($T=1$ and $T=63$), comparing them 
against the unbounded setting ($T=\infty$) of prior work \cite{deletang2024language,mittu:finezip,narashiman:alphazip,valmeekam:llmzip} and showing that thresholding 
enables relevant trade-offs between compression ratio and throughput.

\smallskip\noindent
\textbf{--} \textit{Comprehensive model evaluation.} We benchmark 30 LLMs -- including code-specialized, general-domain, and quantized models, ranging from 120M to 3B 
parameters -- assessing predictive quality, compression performance, and runtime in 
the most extensive study to date. For each model, we analyze parameter count, memory 
footprint, and the impact of code specialization and quantization. Our framework 
reveals broader trends absent from prior work, which considered at most 3 models of 
largely varying sizes: specialized code models achieve the best compression, while 
smaller or quantized models optimize throughput. We also shed new light on scaling 
laws \cite{kaplan2020scaling}: larger models consistently improve compression ratio, but at the cost of  proportionally lower throughput, highlighting a fundamental trade-off between model 
capacity and practical efficiency.

\smallskip\noindent
\textbf{--} \textit{Measured gains.} Our $T$-bounded strategies achieve up to 37\% 
relative compression gain and 40\% faster throughput over the unbounded approach 
($T=\infty$) adopted in prior work~\cite{mittu:finezip,narashiman:alphazip,valmeekam:llmzip}, 
and up to 68\% relative compression gain over general-purpose compressors on reordered 
archives~\cite{boffa:compressibility-swh}, rising to 82\% when compressing individual files.
Although throughput remains lower than that of general-purpose compressors, our pipeline operates on 
individual files (unlike~\cite{boffa:compressibility-swh}), making parallelization trivial and improving random access to compressed files.

\smallskip\noindent
\textbf{--} \textit{Source-code compression.} While prior LLM-based compression literature 
focused exclusively on natural language corpora or images, we are the first to evaluate 
this approach on source code -- a setting of growing importance, as already motivated. 
This context was previously explored by Boffa et al.~\cite{boffa:compressibility-swh} 
using only general-purpose compressors and costly global reorderings over large archives. We show that LLM-based compressors squeeze source code more than natural language, thus suggesting that source code contains regularities that elude general-purpose compressors but are well captured by LLMs.

\smallskip
These findings indicate that LLM-based symbol-ranking combined with thresholding and 
lightweight or quantized models represent a promising path toward practical and 
scalable source-code compression, particularly suited for cold backup scenarios where 
compression ratio takes priority over speed, paving the way for new pipelines, 
LLM-specialized models, and novel strategies to reduce costly LLM-inference calls -- 
challenges that we discuss at the end of the paper.

\section{Theoretical Background}
\label{sec:theo-back}

\textbf{General-purpose compressors.}
We use three general-purpose lossless compressors, both as baselines and as backends in our LLM-based pipelines. They cover two major
and complementary families: \textit{dictionary-based} compression algorithms, represented by \texttt{zstd}, and \textit{block-sorting} compression algorithms, represented by \texttt{bzip2} and \texttt{bzip3}. We exclude statistical encoders (e.g., Arithmetic Coding) since prior LLM-based compression studies~\cite{mittu:finezip,valmeekam:llmzip} show that they substantially increase runtime, reducing overall throughput.

\texttt{zstd} extends the LZ77 paradigm with entropy coding and offers several compression levels, trading speed for compression ratio~\cite{collet2021rfc8878,facebook2025zstd}. The \texttt{bzip} family instead relies on the Burrows--Wheeler Transform, followed by Move-to-Front, Run-Length Encoding, and entropy coding~\cite{burrows:wheeler,bzip2manual}. While \texttt{bzip2} is the classical representative, \texttt{bzip3} introduces several improvements that typically yield stronger compression while preserving competitive decompression speed~\cite{szewczyk2022bzip3}.

\smallskip
\noindent
\textbf{Symbol-Ranking Text Compression.}
The idea of \emph{symbol-ranking text compression} originates from Shannon’s seminal 1951 paper on the entropy of English text~\cite{shannon:prediction-entropy}. In his experiments, human subjects were asked to predict the next character of a sequence, producing a ranked list of symbols according to their likelihood. Shannon formalized three prediction methods: (i) guessing only the most likely symbol ($T=1$), receiving a yes/no feedback and, in case of error, returning also the correct symbol; (ii) continuing to guess the correct symbol by scanning their ranked list (by decreasing likelihood), until it is found, thereby recording its rank position ($T=\infty$); and (iii) a hybrid approach, where guessing proceeds as in the previous method but stops after a fixed threshold $T$ of wrong guesses, at which point the correct symbol is revealed. 

Starting from these three variants introduced by Shannon, we can formally define the symbol ranking method.

\begin{definition}[Symbol Ranking]\label{def:sr}
Let $S=(s_1,\dots,s_n)$ be a string over an alphabet $\Sigma$.
For each position $i$, let $W_i$ denote its \emph{context} (for example, the last $k$ symbols).
A \emph{predictor} associates with $W_i$ a probability distribution over $\Sigma$, yielding an ordered list $\Sigma_{W_i} = (\sigma_0,\dots,\sigma_{|\Sigma|-1})$, where $\mathbb{P}(\sigma_0 \mid W_i) \geq \dots \geq \mathbb{P}(\sigma_{|\Sigma|-1}\mid W_i)$.
The \emph{rank} of the correct next symbol $s_i$ is then defined as
\begin{equation}\label{eqn1}
\text{Rank}(s_i \mid W_i) = j \quad \iff \quad s_i = \sigma_j.    
\end{equation}
The \emph{symbol ranking} transformation maps $S$ into a sequence of integers (ranks) $R=(r_1,\dots,r_n)$, where $r_i=\text{Rank}(s_i\mid W_i)$.
\end{definition}

\begin{definition}[Symbol Ranking with Exceptions]\label{def:sre}
Let $S$ and $W_i$ be as above, and let a predictor associate with each $W_i$ a probability distribution over $\Sigma$, yielding the ordered list $\Sigma_{W_i}$ as in Definition~\ref{def:sr}.
Fix a threshold $T \in \mathbb{N}$. For each position $i$, the output symbol $r_i$ is
\begin{equation}
r_i =
\begin{cases}
\text{Rank}(s_i \mid W_i), & \text{if } \text{Rank}(s_i \mid W_i) \leq T, \\
\text{ESC}, & \text{otherwise,}
\end{cases}    
\end{equation}
where $\text{Rank}(s_i \mid W_i)$ is defined as in Equation~\eqref{eqn1}, and \texttt{ESC} is an escape symbol.
In the case of \texttt{ESC}, the true symbol $s_i$ is stored in a side list $E$.
The output of \emph{symbol ranking with exceptions} is thus the pair $(R,E)$, where $R$ is the sequence of ranks/escapes and $E$ is the list of (symbol) exceptions.
\end{definition}

\noindent
Shannon’s three methods derive from the definitions: the first to Definition~\ref{def:sre} with $T=1$, the second to Definition~\ref{def:sr} (equivalently, Definition~\ref{def:sre} with $T=\infty$), and the third to Definition~\ref{def:sre} with any finite $T>1$.


It is important to stress that symbol ranking is not a compression algorithm per s\'e, but a \emph{transform} producing a sequence of integers with a skewed distribution. These can then be efficiently encoded using general-purpose compressors (e.g., \texttt{zstd}, \texttt{bzip2}, or \texttt{bzip3}). The effectiveness of the method ultimately depends on the quality of the predictor. In particular, the more \emph{skewed} the predicted distributions are, the more tokens fall into low ranks, which directly corresponds to lower entropy and thus higher compressibility. It goes without saying that the predictor must be \emph{deterministic} to ensure full reversibility of the compression process, and thus preserve the lossless property in the decompression stage.


\smallskip
\noindent
\textbf{LLM-Based Predictors and Proposed Pipeline.}
The connection between language modeling and compression has long been recognized~\cite{mahoney2000fast,schmidhuber1996sequential}. Recent works have reinforced this link, showing that accurate next-token prediction implies efficient compression. Once text is tokenized -- with tokens corresponding to the symbols to be compressed -- LLMs can generate {\em "precise"} probability distributions over the next token given a context, making them ideal predictors for symbol-ranking compression. The key advantage of LLMs resides in the fact that, thanks to neural-network processing and the more recent Attention Mechanism~\cite{vaswani2017attention}, the {\em $k$-long context} used to estimate symbol probabilities is not matched {\em exactly}, as in PPM-like approaches (cfr. \cite{cleary1997unbounded,cleary1984ppm,ferragina:pearls,lelewer1991streamlining,moffat1990ppm}), but may be {\em fuzzily and weightly matched}, thus offering a much higher precision in the estimate. 

In the Introduction, we discussed early neural approaches.
All these works -- whether they explicitly adopt a symbol-ranking pipeline \cite{mittu:finezip,narashiman:alphazip,valmeekam:llmzip} or feed predicted 
probabilities into a statistical compressor \cite{deletang2024language,mittu:finezip,valmeekam:llmzip} -- rely solely on the approach described 
in Definition~\ref{def:sr} (i.e., $T=\infty$): no threshold is applied to separate 
high-confidence predictions from rare symbols, leaving unexplored the impact of bounding $T$ on both compression ratio and throughput. These studies consistently show that 
LLM-based predictors outperform general-purpose compressors in compression ratio, but still 
suffer from extremely low throughput. Furthermore, most of them evaluate only a single 
model, and even the most systematic study~\cite{deletang2024language} considers just 
three models with largely varying sizes, without systematically analyzing which 
architectural or training factors most influence compression performance across models 
of comparable scale. Finally, while \cite{deletang2024language} extends compression 
experiments to images and video, all these works focus exclusively on classical text 
corpora, leaving source code -- a setting of growing importance, as motivated in the 
Introduction -- entirely unexplored.


\smallskip
\noindent
\textbf{Our proposal}. We address the limitations of prior work along three 
complementary directions. First, we extend the symbol-ranking framework by 
implementing all three of Shannon's methods (Definitions~\ref{def:sr} 
and~\ref{def:sre}), introducing for the first time a thresholded pipeline ($T$-bounded). 
Second, we benchmark 30 models across architectures and quantization levels on source 
code -- the first large-scale LLM-based compression study on this domain -- advocating 
for compression as a task-agnostic lens to compare foundation models. Third, we show 
that LLMs capture source-code regularities beyond the reach of general-purpose compressors, 
achieving compression ratios surpassing those on natural language.

Methodologically, the first pipeline reproduces the unbounded setting ($T=\infty$) of 
prior work \cite{deletang2024language,mittu:finezip,narashiman:alphazip,valmeekam:llmzip}; the second -- introduced here -- applies a rank threshold $T$, 
replacing out-of-threshold tokens with an escape symbol stored in a side list. Both pipelines share a unified architecture optimized for throughput. Each file is 
processed independently in an \emph{offline setting}, making the pipeline 
parallelizable across GPUs. Within each file, we use a \emph{dynamic context window} 
growing up to a fixed maximum $M$; longer windows are split with zero stride, ensuring 
coverage without overlap. Input sequences are processed in batches of size $B$, enabling parallel rank computation across multiple sequences on GPU.

Finally, the ranked and escaped sequences are compressed using general-purpose lossless compressors -- i.e., \texttt{zstd}, \texttt{bzip2}, and \texttt{bzip3} -- selected for their balance between compression effectiveness and speed. Statistical encoders such as Arithmetic Coding were deliberately excluded, as prior studies \cite{mittu:finezip,valmeekam:llmzip} showed that their runtime penalties outweigh the modest compression gains.

\section{Experimental Setup}
\label{sec:exp-setup}

All experiments were conducted on an NVIDIA DGX H100 server with 8×H100 GPUs (80 GB each), dual Intel Xeon Platinum CPUs, and 2 TiB of RAM, running Ubuntu 22.04. The implementation was developed in Python/PyTorch, with pretrained models accessed via the Hugging Face ecosystem. This setup enabled medium/large-scale LLMs to be tested efficiently under realistic conditions.

\smallskip
\noindent
\textbf{Datasets.} 
We conduct our experiments on \emph{source-code corpora} derived from the Stack-Edu collection~\cite{allal2025smollm2,huggingfaceTB_stackedu}, curated from The Stack v2~\cite{bigcode2024stackv2,lozhkov2024starcoder}, namely the StarCoder2 training corpus. Stack-Edu provides \emph{SWH identifiers} (SWHIDs) enabling us to fetch the corresponding files directly from the Software Heritage archive. This choice is motivated by three main reasons: (i) it ensures that our experiments are conducted on code that already forms part of the Software Heritage Archive, offering a realistic scenario aligned with our goal of designing compression algorithms for this infrastructure; (ii) it guarantees high-quality code aligned with the training corpora of modern LLMs; and (iii) it provides full reproducibility, since the data are publicly accessible.

\smallskip
\noindent
\textbf{Two-phase evaluation strategy.} To avoid long experiments, we adopt a strategy based on a proper dataset design (see Section~\ref{sec:exp-results}), as described below: 

\smallskip    
\noindent
\textbf{--} In \textit{Phase One} (exploratory analysis), we construct a \(\sim 60\)~MB mixed-language dataset by uniformly sampling \(\sim 10\)~MB of source code for each of six representative programming languages: Python, C, C++, C\#, Java, and JavaScript. This smaller dataset allows us to evaluate all 30 LLMs in terms of both predictive accuracy and runtime efficiency, without the computational burden of large-scale experiments. The goal is to identify promising models while discarding those that show poor accuracy or excessive runtime. 

\smallskip    
\noindent
\textbf{--} In \textit{Phase Two} (in-depth analysis), we focus only on the most promising models and therefore build larger, per-language datasets of approximately \(\sim 100\)~MB each, for the same six programming languages used in Phase One, amounting to nearly \(\sim 600\)~MB in total. This design enables a detailed language-specific evaluation of the complete compression pipeline. To ensure comparability across datasets, the average file size was constrained to approximately 3KB per file.

\begin{table}[t]
\centering
\scriptsize
\setlength{\tabcolsep}{2.2pt}
\renewcommand{\arraystretch}{1.08}
\begin{tabular}{l|c|c|c||l|c|c|c}
\hline
\textbf{Model} & \textbf{Par.} & \textbf{Qnt.} & \textbf{Code}
&
\textbf{Model} & \textbf{Par.} & \textbf{Qnt.} & \textbf{Code} \\
\hline
\hline

\textbf{StarCoder2-3B (4bit)} & 3B & Yes & Yes
&

\textbf{Llama-3.2-1B (4bit)} & 1.23B & Yes & No \\

StarCoder2-3B (8bit) & 3B & Yes & Yes
&

Llama-3.2-1B & 1.23B & No & No \\

\textbf{StarCoder2-3B (bf16)} & 3B & No & Yes
&
Llama-3.2-3B (4bit) & 3.21B & Yes & No \\

StarCoder2-3B (fp32) & 3B & No & Yes
&
Llama-3.2-3B & 3.21B & No & No \\

\cline{1-4}

DeepSeekCoder-1.3B & 1.3B & No & Yes
&
Llama-3.1-8B (4bit) & 8B & Yes & No \\

\textbf{DeepSeekCoder-1.3B AWQ }& 1.3B & Yes & Yes
&
Llama-3.1-8B & 8B & No & No \\
\cline{1-4}\cline{5-8}

CodeGemma-2B & 2.5B & No & Yes
&
Mistral-7B & 7B & No & No \\

\textbf{CodeGemma-2B AWQ} & 2.5B & Yes & Yes
&
Mistral-7B (4bit) & 7B & Yes & No \\
\cline{1-4}\cline{5-8}

Gemma-2-2B & 2.6B & No & No
&
Qwen3-1.7B & 1.7B & No & No \\
\cline{5-8}

\textbf{Gemma-2-2B (4bit)} & 2.6B & Yes & No
&
\textbf{UniXcoder-base} & 0.13B & No & Yes \\
\cline{1-4}\cline{5-8}

SmolLM3-3B & 3B & No & No
&
CodeT5-base & 0.22B & No & Yes \\
\cline{1-4}\cline{5-8}

Granite-2B & 2B & No & No
&
GPT-2 small & 0.124B & No & No \\

Granite-Code-3B & 3B & No & Yes
&
GPT-2 medium & 0.355B & No & No \\

\textbf{Granite-Code-3B (4bit)} & 3B & Yes & Yes
&
GPT-2 large & 0.774B & No & No \\
\cline{1-4}

\textbf{Phi-2} & 2.7B & No & No
&
GPT-2 xl & 1.5B & No & No \\
\cline{1-4}\cline{5-8}

\hline
\end{tabular}
\vspace{0.6em}
\caption{Evaluated models. {\em Qnt} denotes quantized models; {\em Code} denotes code-specialized models. Bold entries denote those selected for the in-depth evaluation.}
\label{tab:model_summary}
\vspace{-3em}
\end{table}

\smallskip
\noindent
\textbf{Models.} 
The choice of prediction model is critical, as it directly affects both rank accuracy and runtime performance.
Except for \texttt{CodeT5}, all models considered are of the decoder-only type. Decoder-only architectures are naturally suited to our task, as they are designed for autoregressive generation: given a context, they estimate the probability distribution of the next token -- the same required for rank computation. Encoder-only and encoder–decoder models are less effective for prediction and were thus only marginally included.

Most tested models fall in the small/mid-scale range (1–3B parameters), reflecting 
practical runtime constraints. Two larger representatives, \texttt{LLaMA-3.1-8B} and 
\texttt{Mistral-7B-v0.3}, are included to cover widely adopted LLM families. Overall, 
we evaluate 30 models, many considered here for the first time in a lossless 
compression setting. A schematic summary is provided in Table~\ref{tab:model_summary}, 
reporting parameter count and code specialization or quantization. 
For several families, multiple variants are included (e.g., different precisions or 
code-specific fine-tuning), allowing us to assess how these design choices affect both 
compression performance and the broader predictive capabilities of the models.

\section{Experimental Results}
\label{sec:exp-results}

In this section, we follow the two-phase evaluation strategy mentioned above. 

\subsection{Phase One (Exploratory Analysis)}

In this first phase, we evaluate all 30 LLMs on a 60~MB multilingual code dataset 
(see Section~\ref{sec:exp-setup}), focusing on predictive accuracy -- captured by the 
rank distribution -- and throughput, without performing actual compression. All models 
use a unified pipeline ($T=\infty$, Definition~\ref{def:sr}) with fixed $B=32$ and 
$M=512$, selected through preliminary experiments as the configuration maximizing GPU 
efficiency within memory constraints.

\begin{figure}[t]
    \centering
    \begin{minipage}{0.45\linewidth}
        \centering
        \includegraphics[width=\linewidth]{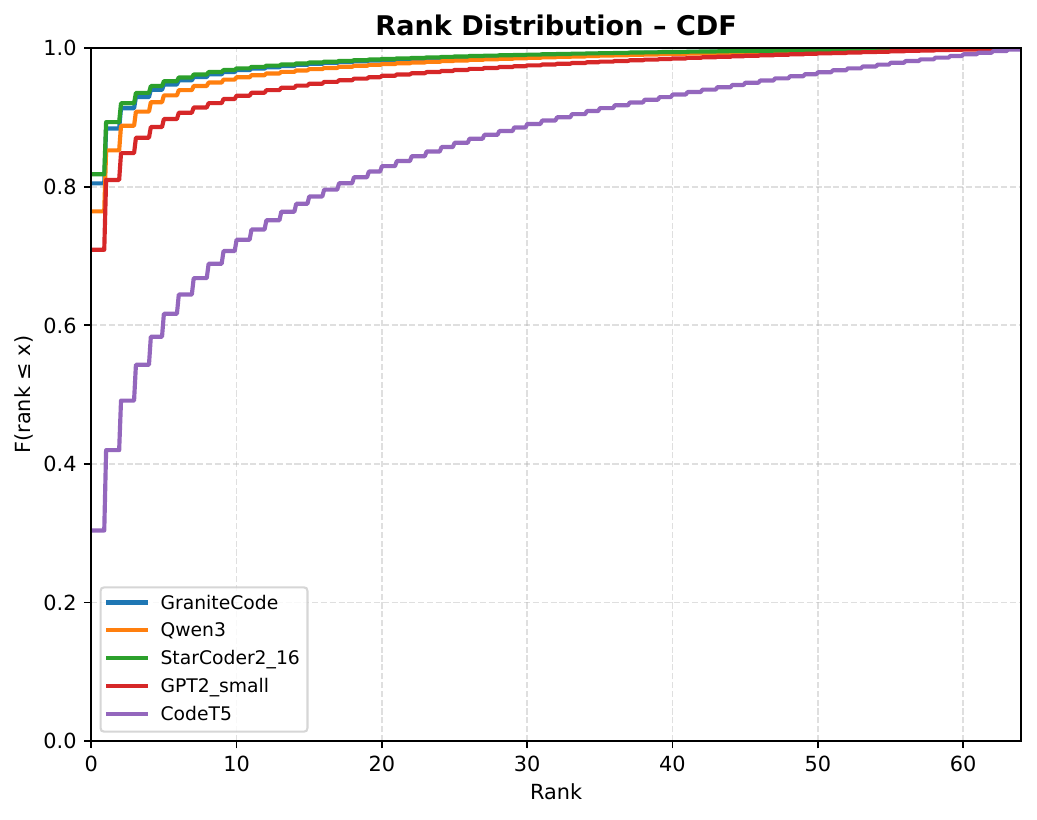}
        \subcaption*{(a) Rank prediction CDF.}
    \end{minipage}
    \hfill
    \begin{minipage}{0.54\linewidth}
        \centering
        \includegraphics[width=\linewidth]{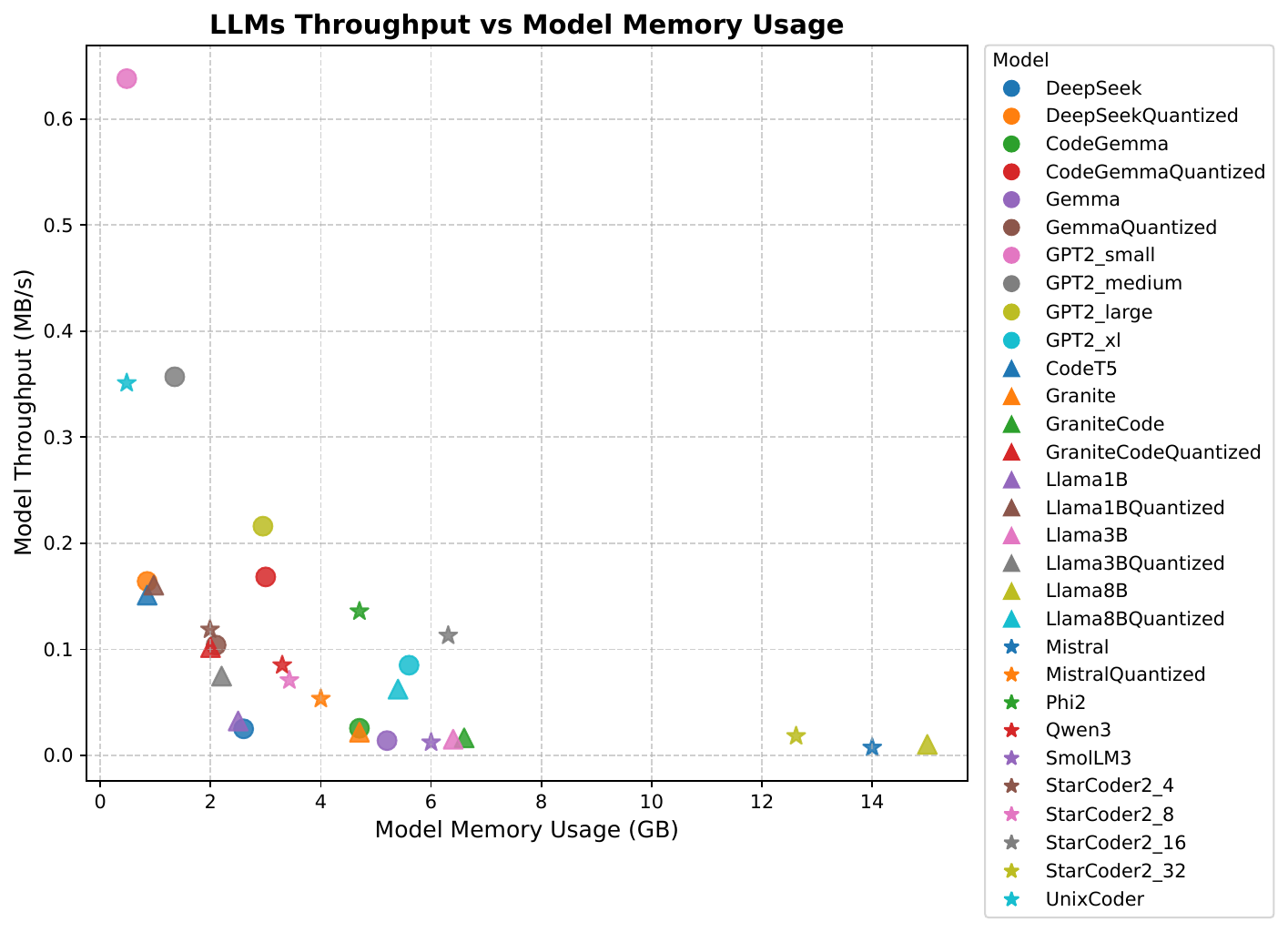}
        \subcaption*{(b) Throughput vs. memory.}
    \end{minipage}
    \caption{Exploratory analysis. (a) CDF of rank predictions for models, highlighting the weak predictive accuracy of \texttt{CodeT5}. (b) Throughput versus memory usage, showing the inverse relation between model size and inference speed.}
    \label{fig:exp1_predictors}
    \vspace{-1.5em}
\end{figure}

\smallskip
\noindent \textbf{Prediction Results.}
From a predictive standpoint, most models exhibit highly skewed rank distributions: rank~0 accounts for about 70–80\% of occurrences, while the first 63 ranks cover 95–98\%. In particular, code-specialized LLMs such as \texttt{StarCoder2}, \texttt{DeepSeekCoder}, \texttt{GraniteCode}, and \texttt{CodeGemma} achieve the skewest rank profiles. Among all evaluated models, two clear outliers emerge: \texttt{CodeT5}, whose encoder–decoder architecture results in very poor rank accuracy (cfr. Figure~\ref{fig:exp1_predictors}(a)), and \texttt{UniXcoder} together with the \texttt{GPT-2} family, which are the only other models falling below the ranges reported above (i.e., 70–80\% and 95–98\%). In particular, the \texttt{GPT-2} family exhibits poor results across all evaluated metrics (e.g., number of tokens, mean, variance), whereas \texttt{UniXcoder} maintains values comparable to those of other models. Consequently, \texttt{CodeT5} and all \texttt{GPT-2} variants are excluded from the following in-depth experimental Phase Two, while \texttt{UniXcoder} is retained as a lightweight baseline.

\smallskip
\noindent \textbf{Time Results.}
Execution-time measurements confirm that throughput constitutes the main bottleneck. Even the fastest models reach at most $\sim$0.6~MB/s, while the majority cluster around or below 0.1~MB/s. A negative correlation emerges between GPU memory footprint and throughput (Figures~\ref{fig:exp1_predictors}(b)): larger models, such as \texttt{Llama-8B} and \texttt{Mistral}, exhibit slower processing rates (i.e., they occur at the rightmost corner). To balance predictive accuracy and speed, we define a selection threshold of 0.1~MB/s and retain only those models that satisfy this throughput constraint and the predictive-quality filter discussed above.


\smallskip
\noindent \textbf{Conclusions.}
The first-phase analysis sheds light on scaling laws \cite{kaplan2020scaling}: larger models improve prediction quality, but reduce throughput, exposing a capacity--efficiency trade-off. 
The nine selected LLMs -- highlighted in bold in Table~\ref{tab:model_summary} -- are 
predominantly quantized, with \texttt{StarCoder2-3B (bf16)} as the 
non-quantized exception. They form the basis of the second experimental phase, where 
we evaluate the full compression pipeline and the proposed thresholded strategies.

\subsection{Phase Two (In-Depth Analysis)}

In this second phase, we move to a complete compression evaluation -- focusing on 
compression ratio and throughput -- across six language-specific datasets, each 
containing 100~MB (see Section~\ref{sec:exp-setup}). We first benchmark general-purpose 
compressors (\texttt{zstd}, \texttt{bzip2}, \texttt{bzip3}) directly on the datasets 
as baselines, then integrate them within the two symbol-ranking pipelines of 
Section~\ref{sec:theo-back}.

\smallskip
\noindent \textbf{Methodology.}
As baselines, we apply \texttt{zstd}, \texttt{bzip2}, and \texttt{bzip3} without LLM involvement, under both \textit{fast} and \textit{high-compression} settings. 
Each compressor is tested in two configurations: (i) files compressed independently; (ii) files reordered by inverted path names and aggregated into a \texttt{.tar} archive, following~\cite{boffa:compressibility-swh}.

For the symbol-ranking pipelines, we keep the same configuration used in Phase One, namely $B=32$ and $M=512$. 
The known pipeline (Definition~\ref{def:sr}, $T=\infty$) records each token's exact rank as a 32-bit integer and compresses the resulting stream with the baseline compressors. 
The new pipeline (Definition~\ref{def:sre}) instantiates Shannon's first method with $T=1$ and his third method with $T \in \{3, 7, 15, 31, 63, 127, 255\}$. 
Ranks are encoded in $\lceil \log_2(T+1) \rceil$ bits, while exceptions are stored separately as 32-bit integers; both streams are then compressed with the same baseline compressors. 
Our experiments identify $T=63$ as the best configuration for Shannon's third method, which we adopt as reference.

\smallskip
\noindent \textbf{Results on General-Purpose Compression Baselines.} Unless otherwise stated, baseline results refer to the archive-based configuration, which achieves the best compression. As expected, \texttt{zstd-3} delivers the highest throughput ($\approx$40 MB/s) but also the worst compression, while higher compression level (i.e., \texttt{zstd-12} and \texttt{zstd-22}) improves compression ratio at the cost of speed. \texttt{bzip2} shows nearly identical throughput across the various compression levels, with minor gains in compression for \texttt{bzip2-9}. In contrast, \texttt{bzip3} offers the best speed–compression trade-off and is therefore adopted as the strongest baseline. 

\smallskip
\noindent \textbf{LLM Pipeline Comparison.}
Across all pipelines, models, and datasets, the execution-time breakdown shows that \emph{rank computation} (i.e., the LLM forward pass), together with tokenization, is the dominant bottleneck, accounting for more than 99\% of the total runtime. As a consequence, throughput is determined by the choice of the model rather than by the downstream compressor, and we can focus on the configurations that achieve the best compression.

Our second pipeline, which introduces a rank threshold $T$, improves over the 
conventional unbounded approach of the first pipeline ($T=\infty$). With $T=1$ (Shannon's first method), 
throughput improves by up to 40\%, at the cost of a small compression loss. In 
contrast, $T=63$ (Shannon's third method) achieves both higher throughput and better 
compression than $T=\infty$. This is due to two factors: (i) more than 95\% of tokens 
fall within the top-63 predictions, keeping the exception list small, and (ii) ranks 
require only 6 bits instead of 32. Overall, the $T=63$ pipeline yields up to 37\% 
additional compression gain and up to 40\% throughput improvement over $T=\infty$ across configurations -- the latter matching $T=1$, since extracting top-1 and 
top-63 predictions is practically time equivalent. These results identify $T=63$, 
combined with \texttt{bzip3-default} or \texttt{zstd-22}, as the most effective 
LLM-based configuration, which we adopt for the final comparison against general-purpose 
compressors.

From the model perspective, consistent patterns emerge across all configurations. 
As expected, the \texttt{StarCoder} family achieves the strongest compression, while the smallest 
model, \texttt{UniXcoder}, maximizes throughput at the cost of weaker compression. 
Models around 1B parameters -- in particular \texttt{Llama3.2-1B} and \texttt{DeepSeekCoder} -- offer the best speed--compression trade-off; remarkably 
without requiring code specialization in the case of \texttt{Llama3.2-1B}.

\begin{figure}[t] 
    \centering 
    \includegraphics[width=1\linewidth]{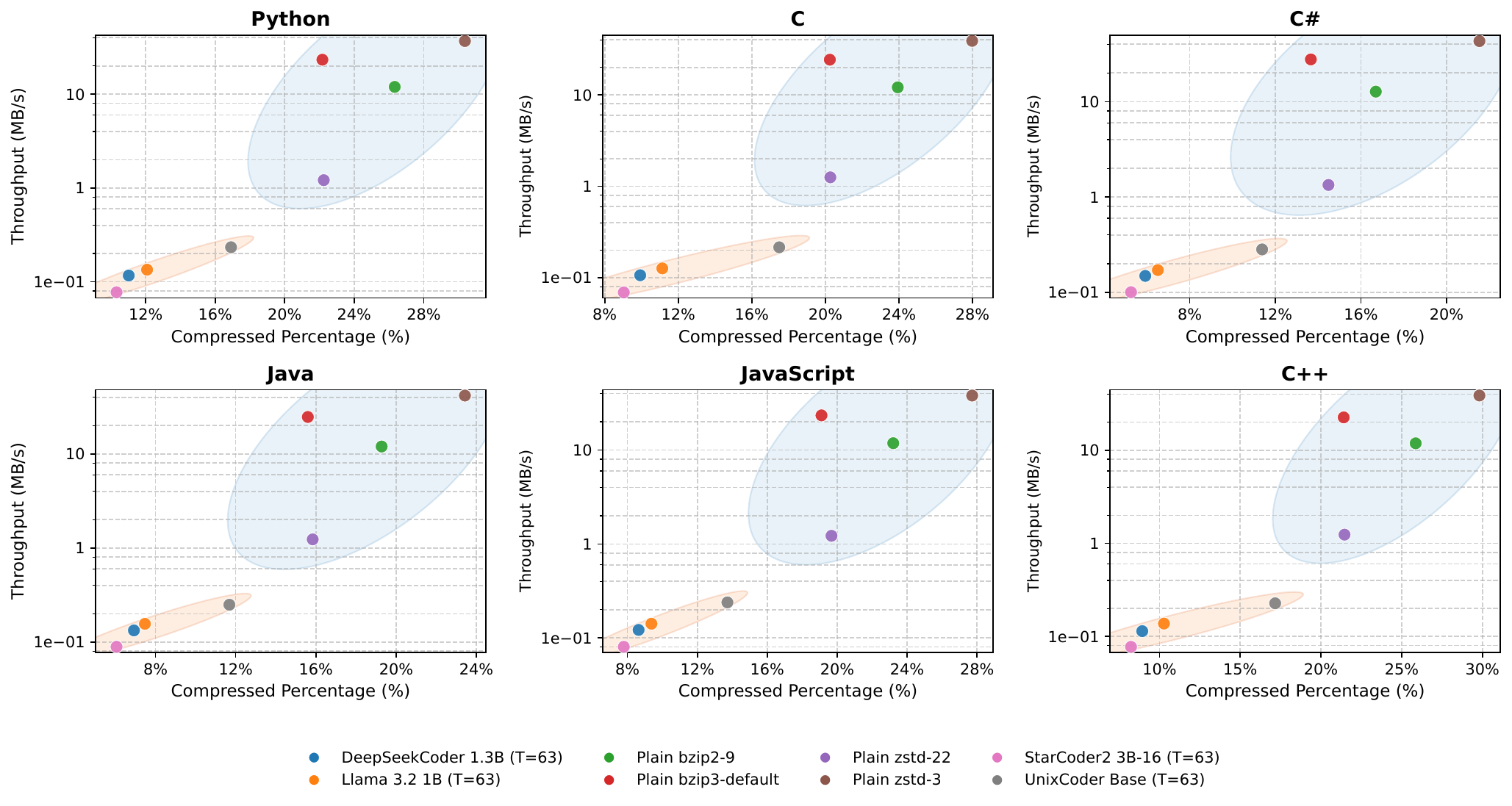} 
    \caption{Throughput (log scale) versus compressed percentage for the best LLM-based pipeline ($T=63$ with \texttt{bzip3-default}) and general-purpose lossless compressors (\texttt{zstd-22}, \texttt{zstd-3}, \texttt{bzip-9}, \texttt{bzip3}) in the tar-based setting.} 
    \label{fig:final_comparison} 
    \vspace{-2em}
\end{figure}

\smallskip
\noindent \textbf{LLM Pipelines vs. General-Purpose Compressors.}
Across all experiments, LLM pipelines compress better than general-purpose compressors, but with significantly lower throughput. Figure~\ref{fig:final_comparison} illustrates this trade-off by comparing our best configuration -- the $T=63$ pipeline with \texttt{bzip3-default} and representative LLMs -- against strong general-purpose compressors in the tar-based setting. In all datasets, the two families occupy distinct regions in the throughput--compression plane: general-purpose compressors cluster in the \emph{upper-right}, offering very high throughput but weaker compression, whereas LLM-based pipelines lie in the \emph{lower-left}, achieving better compression at cost of one--two orders of magnitude lower speed.

Quantitatively, the best LLM configuration reduces compressed size by up to 68\% compared to the strongest baselines. This improvement is significant because our LLM-based method operates on single files (or even partial files, depending on the context window), which makes parallelization trivial. In contrast, general-purpose compressors exploit inter-file similarities by first sorting them (by reversed filename, as in \cite{boffa:compressibility-swh}) and applying them with large context windows: hence they incur the cost of computing those reorderings for large source-code collections, such as in Software Heritage. When general-purpose compressors are instead applied to individual files, the best LLM configuration reduces compressed size by up to 82\%. 
Notably, under both baselines, even \texttt{UniXcoder}, which yields the weakest compression among our selected LLMs, still outperforms the strongest general-purpose compressors (see Figures \ref{fig:final_comparison}). The remaining model-level trends are consistent with the previous analysis.

As far as the decompression throughput is concerned, we mention that it is slower than the compression one. This indicates that current LLM-based symbol-ranking schemes still require throughput improvements before becoming competitive for real-world applications. Nonetheless, our results point to a promising direction, especially when combined with lighter architectures or new schemes that reduce costly LLM predictions, also considering that our pipelines can be parallelized across batches. Further future directions are discussed in Section~\ref{sec:conclusions}.

\smallskip
\noindent \textbf{Text Compression vs. Source Code Compression.}
As a final comparison, Table~\ref{tab:compression_gain_related} contrasts the gains obtained by LLM-based compressors on English text and source code. 
For text, we use the raw compression results reported by Del\'etang et al.~\cite{deletang2024language} on enwik9 with Transformer 800K, Chinchilla-1B, and LLaMA~2 (see Table 1 of \cite{deletang2024language}), and normalize them against \texttt{zstd-22} for a fair comparison. 
For source code, we report our average gains over the six language-specific datasets using \texttt{UniXcoder}, \texttt{Llama3.2-1B}, and \texttt{DeepSeekCoder}, again normalized against \texttt{zstd-22}. 
We consider the two scenarios discussed above: files compressed individually, or reordered by inverted path names and packed with \texttt{tar}~\cite{boffa:compressibility-swh}.


It is easy to observe that LLM-based compression yields larger gains on source code than on English text. It is striking that at comparable model scales, source-code gains are consistently higher: \texttt{DeepSeekCoder} nearly matches the gain of LLaMA~2 despite being much smaller, and the advantage becomes even larger when source-code files are compressed individually. 
These results suggest that source code exposes structural regularities that are well captured by LLMs but less effectively exploited by general-purpose compressors such as \texttt{zstd-22}.

\begin{table}[t]
\centering
\scriptsize
\setlength{\tabcolsep}{2.5pt}
\renewcommand{\arraystretch}{1.12}
\begin{tabular}{@{}lllcccc@{}}
\toprule
\textbf{Work} & \textbf{Data} & \textbf{Method} & \textbf{Model} & \textbf{Params} & \textbf{Mode} & \textbf{Gain} \\
\midrule
\multirow{3}{*}{Del\'etang et al.~\cite{deletang2024language}} 
& \multirow{3}{*}{\begin{tabular}{@{}c@{}}English\\text\end{tabular}}
& \multirow{3}{*}{\begin{tabular}{@{}c@{}}Arithmetic\\coding\end{tabular}}
& Transformer-like 
& 800K
& one file 
& $0.99\times$ \\

& 
& 
& Chinchilla 
& 1B
& one file
& $1.90\times$ \\

& 
& 
& Llama2 
& 7B
& one file
& $2.59\times$ \\

\midrule

\multirow{6}{*}{This work} 
& \multirow{6}{*}{\begin{tabular}{@{}c@{}}Source\\code\end{tabular}}
& \multirow{6}{*}{\begin{tabular}{@{}c@{}}Symbol\\ranking\\(\texttt{zstd-22})\end{tabular}}
& UniXcoder 
& 0.13B
& collection 
& $1.89\times$ \\

& 
& 
& Llama-3.2-4bit 
& 1B
& collection 
& $3.04\times$ \\

& 
& 
& DeepSeekCoder-AWQ 
& 1.3B
& collection 
& $3.29\times$ \\

\cmidrule(lr){4-7}

& 
& 
& UniXcoder 
& 0.13B
& sorted coll. 
& $1.18\times$ \\

& 
& 
& Llama-3.2-4bit 
& 1B
& sorted coll. 
& $1.90\times$ \\

& 
& 
& DeepSeekCoder-AWQ 
& 1.3B
& sorted coll. 
& $2.35\times$ \\

\bottomrule
\end{tabular}
\vspace{0.6em}
\caption{Compression gains of LLM-based compressors on English text~\cite{deletang2024language} 
and source code (this work), relative to \texttt{zstd-22}. Our results are averaged 
across six programming languages.}
\label{tab:compression_gain_related}
\vspace{-3em}
\end{table}

\section{Conclusions and Future Works}
\label{sec:conclusions}

Building on Shannon’s symbol-ranking framework, we conducted the first systematic study of LLM-based compression on source-code corpora, and proposed new approaches that yield relative compression gains up to 37\% and throughput improvements up to 40\% over existing LLM-based methods, while also surpassing general-purpose compressors by up to 82\% in relative compression gain, yet remaining slower. Our results show that LLM-based compressors are especially effective on source code, achieving larger gains than on natural language.

\smallskip
\noindent
\textbf{Future Works}. Despite these advances, throughput remains the limitation of LLM-based compressors. Future work can be divided into two directions. The first concerns pipeline refinements, aimed at reducing the cost of LLM inference: (i) \emph{parsing-based preprocessing} (e.g., Prefix-Free Parsing \cite{boucher2019prefixfree}) to eliminate redundancy before rank computation, and thus reducing the number of LLM-inference steps; (ii) \emph{skipping the first tokens}, where prediction is known to be inaccurate; (iii) dedicated handling of \emph{flat probability distributions}, where rank computation offers little benefit; and (iv) \emph{parallel processing} of batches to improve throughput. 

Another direction concerns the extension to {\em larger datasets} and the development of specialized models for compression: (i) \emph{distillation} of large into lightweight models; (ii) \emph{hardware-aware optimizations}, such as bfloat16 inference or quantization. 


%
%
%
%





%
%
\bibliographystyle{splncs04}
\bibliography{references}

\end{document}